\documentstyle[12pt,epsf]{article}

\author{Re'em Sari
\and
Tsvi Piran}
\title{Variability in GRBs - A Clue
}

\def\sles{\lower2pt\hbox{$\buildrel {\scriptstyle <} 
   \over {\scriptstyle\sim}$}}
\def\sgreat{\lower2pt\hbox{$\buildrel {\scriptstyle >} 
   \over {\scriptstyle\sim}$}}

\begin{document}

\maketitle
\begin{abstract}
We show that external shocks cannot produce a variable GRB, unless they are
produced by an extremely narrow jets (angular opening of $\sles 10^{-4}$)
or if only a small fraction of the shell emits the radiation and the
process is very inefficient. Internal shocks can produce the observed
complex temporal structure provided that the source itself is variable. In
this case, the observed temporal structure reflects the activity of the
``inner engine'' that drives the bursts. This sets direct constraints on it.
\end{abstract}

\section{Introduction}

Five years of BATSE's observations with perfect isotropy and paucity of weak
bursts shows that the origin of GRBs is probably cosmological.
Therefore, given the measured flux, GRBs involve immense amount of energy $%
\sim 10^{51}$ergs. The ``compactness problem'' then shows that the observed $%
\gamma $-rays must be emitted by a medium with highly relativistic
velocities having Lorentz factor $\gamma \ge 100$ (Fenimore, Epstein \& Ho
1993, Woods \& Loeb 1995, Piran 1995). While the energy source varies
from one model to another (binary neutron stars merge, failed supernovae or
collapse of magnetic stars) and is relatively speculative, all models of
cosmological GRBs involve a relativistic moving shell which converts its
(kinetic or magnetic) energy to radiation at a large radius. In all these
models the observed radiation does not emerge directly from the ``inner
engine'' that drives the shell, which remains hidden.

Most bursts are highly variable with a variability scale significantly
smaller than the overall duration. Following Fenimore, Madras and Nayakshin
(1996), we use kinematic considerations to constrain different GRB models.
We show that the overall duration, $T$,
reflects directly the length of time that the ``inner engine'' operates and
the observed temporal variability reflects variability in the ``inner
engine''. The only exceptions to this conclusion are if the engine produces
an extremely narrow jet or if GRBs are extremely
inefficient. These considerations also limit the emission radius-the place
where the energy of the shell is converted to radiation, $R_e$, to be
significantly smaller than what was previously thought. The maximal emission
radius is quite close to the minimal radius at which a GRB can be produced
without becoming optically thick. This is also the place where ``internal
shocks'' would naturally take place. Thus, our conclusions are consistent
with the ``internal shock'' scenario. Sufficiently small radii are
impossible in the hydrodynamic version of the ``external shock'' scenario
(and probably in other versions of this scenario as well) .

In section 2 we discuss the angular spreading problem, which is the key to
our discussion. We show in section 3 that in the framework of models in
which the duration of the burst is given by the radius of emission,  all
solutions to the angular spreading problem result in extremely narrow jets or
an extremely low efficiency. In section 4 we discuss models in
which the total duration of the burst corresponds directly to the time that
the ``inner engine'' operates. The internal shock scenario fits this picture.
We show that the hydrodynamic version of the external shock scenario (and most
likely all other versions) is incompatible with these limits.


\section{Angular Spreading}

Special relativistic effects determine the observed duration of the burst
form a relativistic shell. Consider an infinitely thin relativistic shell
with a Lorentz factor $\gamma_e$ (the subscript e is for the emitting
region) and an angular width larger than $\gamma_e^{-1}$. Because of
relativistic beaming an observer can see only a region of size $\gamma_e^{-1}
$. Therefore a shell with an angular size larger than $\gamma_e^{-1}$ can be
considered as spherical. Let $R_e$ be a typical radius characterizing the
emitting region (in the observer frame) such that most of the emission takes
place between $R_e \pm \Delta R_e/2$. The observed duration between the first
photon (emitted at $R_e-\Delta R_e/2$) and last one 
(emitted at $R_e+\Delta R_e/2$) is: 
\begin{equation}
\label{Tradial}T_{radial}\cong \Delta R_e/2\gamma_e ^2c~. 
\end{equation}

Because of radiation beaming an observer sees up to solid angle of $%
\gamma_e^{-1}$ from the line of sight. Two photons emitted at the same time
and radius $R_e$, one on the line of sight and the other at an angle of $%
\gamma_e^{-1}$ away, travel different distances to the observer. The
difference, $R/2\gamma_e ^2$ leads to a delay in the arrival time by
(Ruderman, 1975; Katz, 1994): 
\begin{equation}
\label{Tangular}T_{angular}\cong R_e/2\gamma_e^2c~. 
\end{equation}
Fenimore, Madras and Nayakshin (1996) have shown that the observed pulse
will have a fast rise and a slow decay with FWHM $\sim 0.22 R_e/\gamma_e^2c$.

Comparison of Eqs. \ref{Tradial} and \ref{Tangular} using $\Delta R_e \le R_e
$ reveals that $T_{angular} \ge T_{radial}$. As long as the shell is
spherical on an angular scale larger than $\gamma_e^{-1}$, any temporal
structure that could have risen due to irregularities in the radial
structure of the shell or the material that it encounters will be spread on
a time given by $T_{angular}$. Thus $T_{angular}$ is a lower limit for the
observed temporal variability: $\delta T \ge T_{angular}$.

If the shell has a finite thickness, $\Delta$, (measured in the observer's
rest frame) then the duration of the burst must be longer than $\Delta/c$.
We therefore have: 
\begin{equation}
T= \cases { T_{angular}=R_e/c\gamma_e^2 & if $\Delta < R_e/\gamma_e^2$
{\rm \ \ (Type-I )};
\cr
\Delta/c & otherwise {\rm \ \ \ \ \ \ \ (Type-II)}. \cr }
\end{equation}
It is convenient to classify different GRB models to Type-I and Type-II
according to whether the first or second possibility takes place.

In Type-I models, the burst's duration is determined by the emission radius
and it is independent of $\Delta$. These models include the standard
``external shock model'' (M\'esz\'aros and Rees 1992,1993, Katz 1994, Sari
and Piran 1995) in which the relativistic shell is decelerated on the ISM,
the relativistic magnetic wind model (Usov 1994) in which a magnetic
Poynting flux runs into the ISM, or the scattering of star light by a
relativistic shell (Shemi 1993, Shaviv and Dar 1995).

In Type-II models, the burst's duration is determined by the thickness of
the shell. These models include the ``internal shock model'' (Rees and
M\'esz\'aros 1994, Narayan, Paczy\'nski and Piran 1992, Sari \& Piran 1997),
in which different parts of the shell move with different Lorentz factors
and collide with one another. A magnetic dominated version is given by
Thompson (1994).

The majority of GRBs have a complex temporal structure (e.g. Fishman and
Meegan 1995, Meegan {\it et al.} 1996) with typical variations on a
time-scale, $\delta T$, significantly smaller than the total duration $T$.
We define the ratio $N\equiv T/\delta T$ which is a measure of the
variability and an upper limit for the number of peaks. 
Figure 1 presents a burst of duration $T\sim
75\sec $ and typical peaks of width $\delta T\sles 1\sec $, thus $N\sim100$.
We adopt these as canonical numbers for this letter.

Consider a Type-I model, where $\Delta /c<T_{angular}$ and $T=T_{angular}$.
Angular spreading means that any variability in the emission on a time scale
smaller than $T_{angular}$ is erased unless the spherical symmetry is broken
within angular size smaller than $\gamma _e^{-1}$. Thus a burst produced
from such a shell, in a spherical geometry, must be a smooth single humped
burst with $N=1$ and no temporal structure on a time-scale $\delta T\ll T$.
Put in other words, a shell of a Type-I model, and with angular width larger
than $\gamma _e^{-1}$ cannot produce a variable burst with $N\gg 1$. This is
the angular spreading problem. Fenimore, Madras
and Nayakshin (1996) called this ``the curvature effect''.

On the other hand a Type-II model, contains a thick shell $\Delta
>R_e/\gamma _e^2$, and can produce a variable burst. The variability time
scale, is again limited by $\delta T>T_{angular}$, however $T_{angular}$ can
be shorter than the total duration $T$. The temporal variability can
reflect now radial inhomogeneity of the shell. Since the width, $\Delta$,
is determined by the time that the ``inner engine'' operates, and radial
inhomogeneities in the shell reflects its variability, we find that both the
total duration and the variability time scale reflect those of the source.
This is a remarkable conclusion in view of the fact that the fireball hides
the ``inner engine'' and that it was believed that we would not be able to
obtain any direct information on it.

\section{Angular Variability and Other Caveats}

Thin shells, with $\Delta <R_e/\gamma _e^2$, can produce variable bursts
{\it only} if the opening angle of the emitting region is sufficiently 
small, that is spherical symmetry is broken on scales
 significantly narrower than $\gamma _e^{-1}$. Otherwise the angular spreading
will erase any variability on short time scales.

We begin with estimating the maximal size of an emitting region that can
produce temporal structure of the order of $\delta T=T/N$. Imagine two
points $(r_1,\theta _1)$ and $(r_2,\theta _2)$, $r$ being the distance from
the origin and $\theta $ the angle from the line of sight, that emit
radiation at time $t_1$ and $t_2$ respectively. In principle, one can
carefully choose the emission points $(r_1,\theta _1)$ and $(r_2,\theta _2)$
to produce an arbitrarily narrow pulse. For example one can arrange that the
emitting regions are located on the ellipsoid which is the locus of points
from which photons reach the observer at the same time. However these
ellipsoids are different for different observers and what looks shorter for
a specific observer will look longer to most other observers. The same is
true if we vary the emission time, $t_1$ and $t_2$. Therefore, we assume
that $r_1=r_2=R_e$ and $t_1=t_2$. Consequently, quite generally the
difference in the arrival time between two photons will be: 
\begin{equation}
\label{DeltaT}\delta T\approx \frac{R_e(\theta _2^2-\theta _1^2)}{2c}=\frac{%
R_e\bar \theta \left| \theta _2-\theta _1\right| }c=\frac{R_e\bar \theta
\delta \theta }c~,
\end{equation}
where we have used $\theta _1,\theta _2\ll 1$, $\bar \theta \equiv (\theta
_1+\theta _2)/2$ and $\delta \theta \equiv |\theta _2-\theta _1|$.

Since an observer sees emitting regions up to an angle $\gamma _e^{-1}$ away
from the line of sight $\bar \theta \sim \gamma _e^{-1}$, and the size of
the emitting region $r_s=R_e\left| \theta _2-\theta _1\right| $ is limited
by: 
\begin{equation}
\label{robj}r_s\le \gamma _ec\delta T~.
\end{equation}
The corresponding angular size is: 
\begin{equation}
\label{ang_em}\delta \theta \le {\frac{\gamma _ec\delta T~}{R_e}}={\frac
1{N\gamma _e}}~.
\end{equation}
Note that Fenimore, Madras and Nayakshin (1996) considered only emitting
regions that are directly on the line of sight for which $\bar \theta \sim
\left| \theta _2-\theta _1\right| $ and obtained the limit  $r_s=\gamma _ec
\sqrt{T\delta T}~$ which is larger than our estimate in Eq. \ref{robj}.
However only a small fraction of the emitting regions will be exactly on the
line of sight. Most of the emitting regions will have $\bar \theta \sim
\gamma _e^{-1}$.

The above discussion suggests that one can produce GRBs with $T\approx
T_{radial}\approx R_e/c\gamma _e^2$ and $\delta T=T/N$ if the emitting
regions have angular size smaller than $1/N\gamma _e$. The first idea that
comes to mind is a narrow jet. However, for a typical burst the maximal
opening angle is smaller than $10^{-4}$! Hydrodynamic acceleration can
produce jets with angular width $\gamma _e^{-1}$ or larger. The jets
require another acceleration mechanism. Additionally, either
rapid modulation of the jet or inhomogeneities in the ISM are required 
to produce the temporal variability.
These two options are depicted in Figure 2.

The second possibility is that the shell is relatively ``wide'' (wider than $%
\gamma _e^{-1}$) but the emitting regions are narrow. An example of this
situation is described schematically in Figure 3. This may occur if
either the ISM or the shell itself are very irregular. However the
emitting regions will have a small covering factor, and this situation
is extremely inefficient. The area of the observed part of the shell
is $\pi R_e^2/\gamma _e^2$. To comply with the temporal constraint, the total 
area of the emitting regions is $N\pi r_s^2$.
The ratio between the two, i.e., the fraction of the shell
which emits the radiation is
\begin{equation}
{ N \pi r_s^2  \over \pi R_e^2/\gamma^2 } \le \frac 1{4N}\ll 1,
\end{equation}
where we have used the definition of type-I models, $R_e=2c\gamma _e^2T$,
and Eq. \ref{robj} for the maximal size of the emitting objects. 
This sets an upper limit for the efficiency which is less than $1\%$.

To obtain a high efficiency, i.e., a covering factor of order unity, with
emitting regions of size $r_s$, we must have $\sim 4N^2$ emitting regions. 
But a sum of $4N^2$ peaks each of width $1/N$ of the total duration does not
produce a complex time structure. Instead it produces a smooth time profile
with small variations, of order $1/2N\ll 1$, in the amplitude.

The problem of Type-I models, with shells that are spherical on angular
size of more than $\gamma _e^{-1}$ is fundamental. It does not depend on the
nature of the emitting regions: ISM clouds, star light or fragments of the
shell. This is the case, for example, in the models of Shaviv and Dar (1995)
who consider interaction of a smooth shell with external fragmented medium.
This low efficiency poses a serious energy crisis for most (if not all)
cosmological models. Recall the huge amount of energy, $10^{51}$erg,
observed. It is difficult to imagine sources that can emit considerably
larger amounts of energy, of the order of $10^{53}$erg or more in the form
of a relativistic shell. If such sources exist, it is not clear what will
happen to the rest of the kinetic energy of the shell.
Note that a flux of $10^{53}$ ergs per $10^6$ years pre galaxy,
 of $100$GeV cosmic rays is
comparable to the observed cosmic ray background at that energy.

\section{Type-II Models}

The simplest solution to the angular spreading problem is if the emission
radius is sufficiently small so that angular spreading does not erase
temporal structure with time scale $\delta T$ i.e., $R_e \le 2\gamma_e
^2c\delta T$. This can take place in Type-II models in which the overall
duration is $T=\Delta/c$, longer than $T_{angular}=R_e/2\gamma_e ^2c$.

In this class of models one needs multiple shells to account for the
observed temporal structure. Each shell produces an observed peak of
duration $\delta T$ and the whole complex of shells (whose width is $\Delta$%
) produces a burst that lasts $T=\Delta/c$. The observed temporal structure
will be the longer between the temporal structure of the ``inner engine''
and the angular spreading time.

Thus, unlike previous worries (Piran 1995, M\'esz\'aros 1995), we find that
there is some direct information that we have on the ``inner engine'' of
GRBs. It must be capable of producing the observed complicated temporal
structure. This severely constrains numerous models.

Type-II behavior arises naturally in the internal shock model
(M\`esz\`aros  and Rees 1994, Narayan, Paczy\'nski and Piran 1992) where the
shells are created with variable Lorentz factors and therefore collide with
one another and convert a considerable fraction of their kinetic energy into
internal energy which is radiated (Figure 4). Sari and Piran (1997) have
recently given both an upper limits (above external shocks occurs before the
internal shocks) and a lower limits (below which the flow is optically thick)
for the Lorentz factor of the internal shocks: 
\begin{equation}
100\le \gamma _e\le 1200\left( {\frac{\delta T}{1{\rm sec}}}\right)
^{-1/2}\left( {\frac{\ T}{100{\rm sec}}}\right) ^{1/8}l_{18}^{3/8}.
\end{equation}
where $l\equiv (E/c^2n_1)^{1/3}\sim 10^{18}$cm.
The corresponding radius, $R_e$, is given by 
\begin{equation}
3\times 10^{14}\left( {\frac{\delta T}{1{\rm sec}}}\right) \le R_e\le
4\times 10^{16}\left( {\frac{\ T}{100{\rm sec}}}\right) ^{1/4}l_{18}^{3/4},
\end{equation}
 The lower limits might 
be higher for low values of $\delta T$, the full expression are in Sari
and Piran (1997). 

It is worth while to recall that internal shocks can extract at most half of
the shell's energy, and a  relativistic shell with kinetic energy and
Lorentz factor comparable with the original one is left. If the shell is
surrounded by ISM and collisionless shocks occur the relativistic shell will
dissipate by ``external shocks'' as well, which predicts an additional smooth
burst, with comparable energy. The additional burst, whose time scale and
spectrum depend on model parameters, was not yet observed. An alternative is
that the shell continues to move freely and eventually contribute low energy
cosmic rays of about $10^2-10^4$GeV, depending on the Lorentz factor of the
shell. This flux is about $10^{-2}$ of the observed flux at $10^2$GeV.
It is almost comparable to the observed flux if all particles are above 
$10^4$GeV.

While internal shocks are naturally Type-II, it is interesting to ask
whether external shocks could give rise to Type-II behavior. This would have
been possible if we could have set the parameters of the external shock 
model to satisfy $R_e\le 2\gamma _ec\delta T$. For thin shells the deceleration
radius is given by (M\`es\`zaros and Rees 1992):
\begin{equation}
\label{R_e}R_e=l\gamma ^{-2/3}
\end{equation}
and the observed duration therefor $T=R/\gamma ^2=$ $l\gamma ^{-8/3}$. 
The deceleration is gradual and the Lorentz
factor of the emitting region $\gamma _e$ is similar to the original Lorentz
factor of the shell $\gamma $. It seems that with an arbitrary large Lorentz
factor $\gamma $ we can obtain a small enough deceleration radius $R_e$, as
required by Type-II. However Sari \& Piran (1995) have shown that equation 
\ref{R_e} is valid only for thin shells satisfying $\Delta >l\gamma ^{-8/3}$%
. As $\gamma $ increases above a critical value $\gamma \ge \gamma
_c=(l/\Delta )^{3/8}$ the shell can no longer be considered thin. In this
situation the reverse shock penetrating the shell becomes ultra-relativistic
and the shocked matter moves with Lorentz factor $\gamma _e=\gamma _c<\gamma 
$ independent of the initial Lorentz factor of the shell $\gamma $. The
deceleration radius is now given by $R_e=\Delta ^{1/4}l^{3/4}$ and it is
also independent of the initial Lorentz factor of the shell. The behavior of
the deceleration radius $R_e$ and observed duration as function of the shell
Lorentz factor $\gamma $ is given in Figure 5 for a shell of thickness $%
\Delta =3\times 10^{12}$cm.

\section{Conclusions}

Relativistic motion is essential in GRBs. Relativistic Kinematic arguments
(Fenimore, Madras and Nayakshin 1996) strongly limit GRBs models.
The duration of a GRB is
determined either by the width of the emitting shell, $T=\Delta /c$, or by
the emission radius, $T_{angular}=R_e/\gamma _e^2$. Models in which $%
T_{angular}>\Delta /c$ (type I) can produce only a single hump smooth
bursts. They cannot produce a variable burst. The standard ``external
shock'' model is the classical example of a type I model. Therefore, this
conclusion rules out this scenario. An exception to this conclusions is if
the ``inner engine'' emits an extremely narrow jet (with angular width
smaller than $1/N\gamma _e\sim 10^{-4}$). Such a jet cannot be produced by a
standard fireball in which the matter is accelerated by thermal pressure and
it requires another acceleration mechanism. Alternatively a variable burst
can be produce by irregular shell with angular fluctuations of large
amplitude and small size,or with highly irregular ISM. In this case the
process is extremely inefficient due to low covering factor and the total
energy needed for a GRB is N times larger than the observed energy (of the
order of $10^{53}$ergs).

Type II models does not suffer the angular spreading problem and can produce
the observed temporal structure. The ``internal shock'' model is the
classical example for this type. Note that it is impossible to change the
parameters of an ``external shock'' model so that it will become of type II.
In this case the temporal structure reflects the activity of the ``inner
engine''. The over all duration is the time it operates while the
variability reflects the variability of the source. This is good news
since we have now a direct information on the ``inner engine''. It is also
bad news since only a few known models can produce the observed highly
variable temporal structure observed in GRBs.

\section{Acknowledgments}
We would like to thank Jonathan Katz and Ramesh Narayan for helpful 
discusions. This research was supported in part by a grant from
the US-Israel BSF and by a NASA grant NAG5-1904.

{}

\section*{Figure Captions}

\noindent

Figure 1: A part of burst 2553 of duration $T_{90}=75$sec. The variability 
is on time scale $\sles 1$sec. The variability parameter for this burst is
$N \sim 100$.

Figure 2. Possibilities of creating a variable burst with a very narrow jet
of angular size considerably smaller than $\gamma_e^{-1}$, for which the
angular spreading problem does not exist. The duration of the burst is
determined by the deceleration distance $\Delta R_e$, while the angular time
is assumed small. The variability could now be explain by either variability
in the source which leads to a pulsed jet  (a) or by a uniform jet
interacting with an irregular ISM (b).

Figure 3. An attempt to produce variability in Type-I models by  breaking
the spherical symmetry. A shell with angular size $\gamma^{-1}$ is drawn
(the angular size is highly exaggerated). The spherical symmetry in this
example is broken by the presence of bubbles in the ISM. The relative
angular size of the shell and the bubbles is drawn to scale assuming that a
burst with $N=15$ is to be produced. Consequently $N=15$ bubbles are drawn
(more bubbles will add up to a smooth profile). The fraction of the shell
that will impact these bubbles is small leading to high inefficiency. As $N$
increases the efficiency problem becomes more severe $\sim N^{-1}$.

Figure 4. The internal Shock scenario. The source produces multiple shells
with small fluctuation in the Lorentz factor. These shells catch up with
each other and collide converting some of their kinetic energy to internal
energy. This model is a Type-II one and naturally produces variable bursts.

Figure 5. The external shock problem. The deceleration radius $R_e$ and the
Lorentz factor of the shocked shell $\gamma _e$ as function of the initial
Lorentz factor $\gamma $, for a shell of fixed width $\Delta =3\times
10^{12}cm$. For low values of $\gamma $, the shocked material moves with
Lorentz factor $\gamma _e\sim \gamma $. However as $\gamma $ increases the
reverse shock becomes relativistic reducing significantly the Lorentz factor 
$\gamma _e<\gamma $. This phenomena prevents the ``external shock model''
for being Type-II.

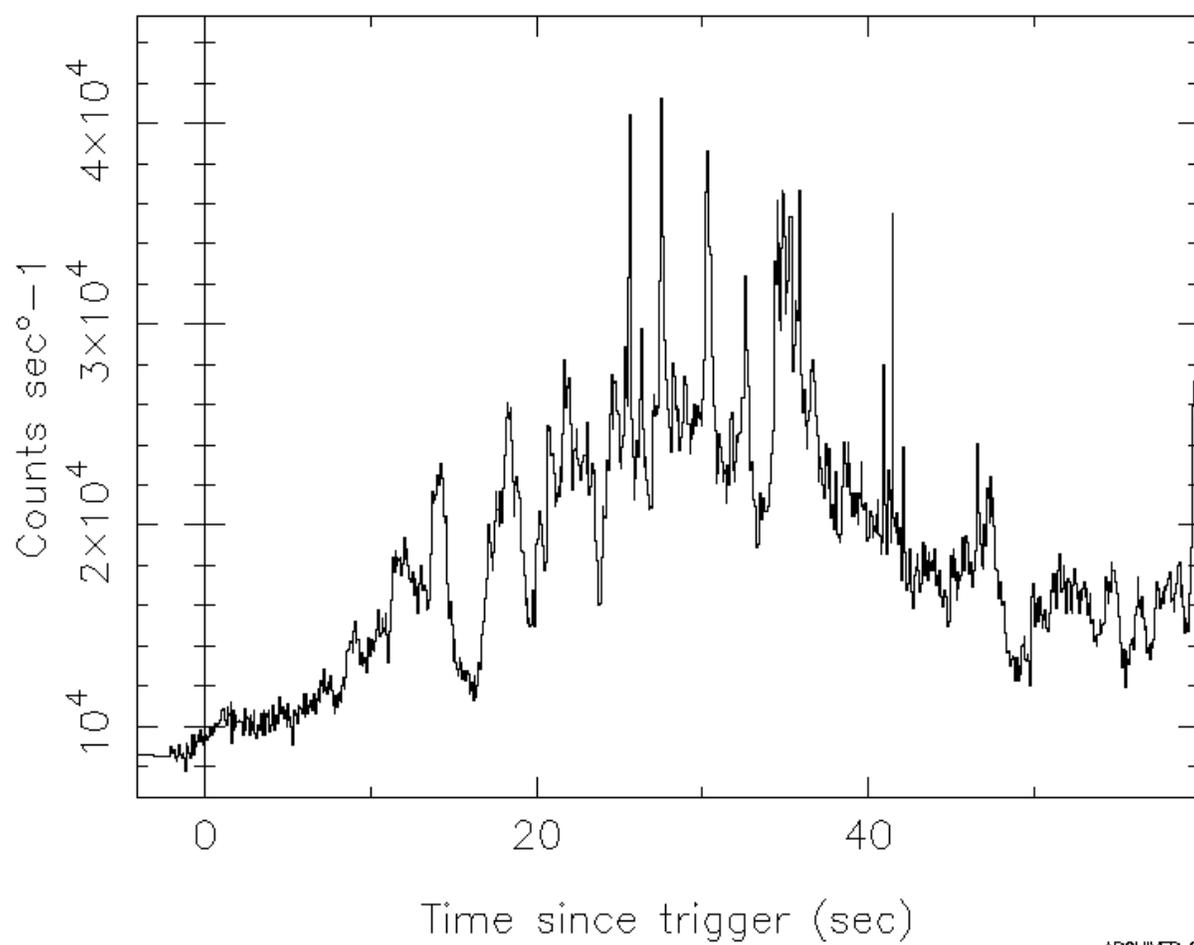
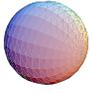
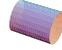
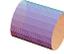
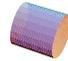
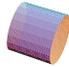
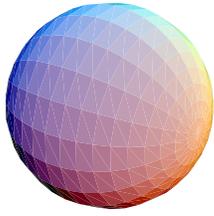
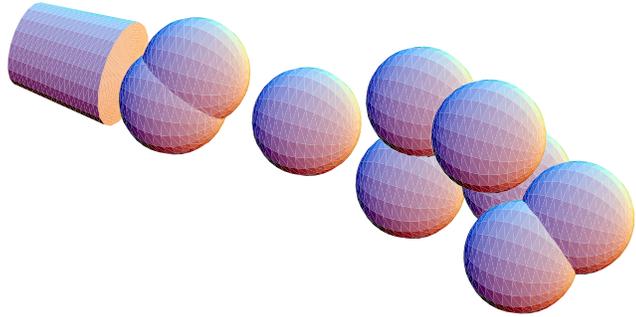
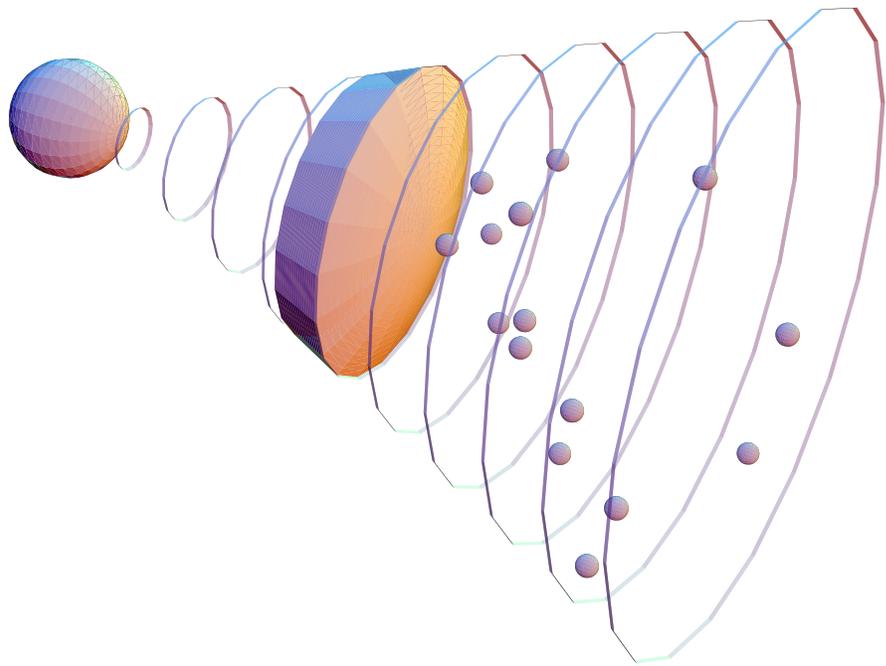
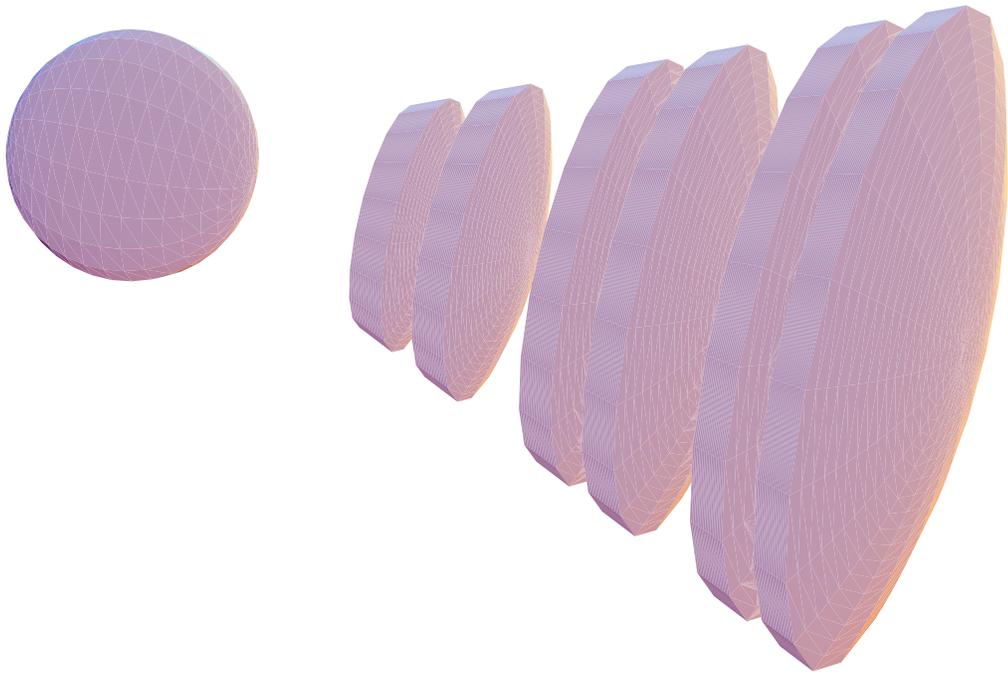
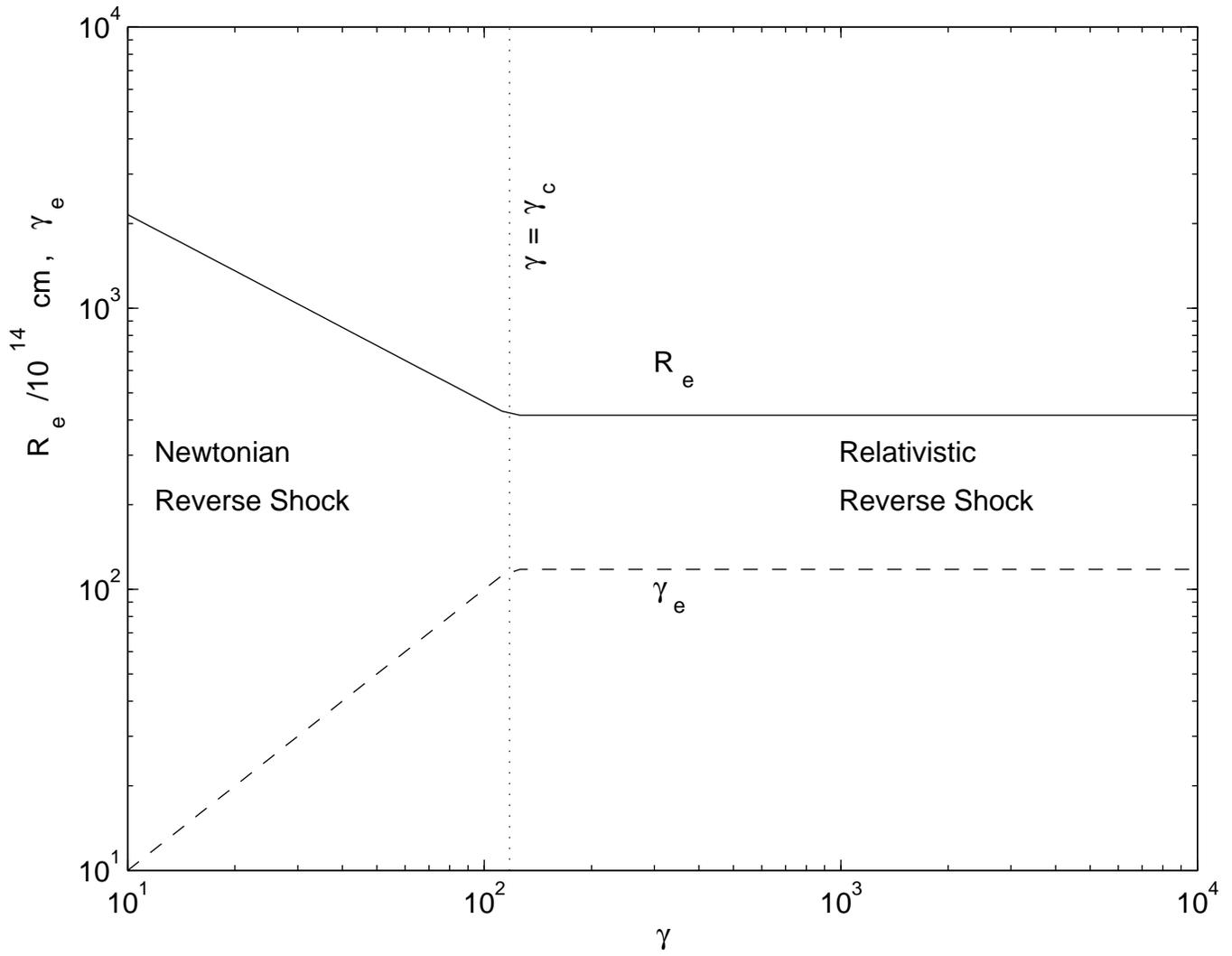
\end{document}